\documentclass[11pt,a4paper]{article}
\usepackage{color}
\usepackage{amsfonts}
\usepackage{amssymb}
\usepackage{graphicx}
\usepackage{amsmath}
\usepackage{enumerate}
\usepackage{color}

\RequirePackage{mathrsfs} \RequirePackage[sc]{mathpazo}
\RequirePackage{wasysym} \RequirePackage{setspace} \textheight=650pt
\newcommand{\be}{\begin{equation}}
\newcommand{\ee}{\end{equation}}
\newcommand{\bea}{\begin{eqnarray}}
\newcommand{\eea}{\end{eqnarray}}
\begin{document}

\title{  { \begin{flushright}
{\normalsize\small } \end{flushright} }\bf
    F-theory and  $N=1$   Quivers   from Polyvalent Geometry }
\author{  Adil Belhaj$^1$ and  Moulay  Brahim  Sedra$^{2}$ \hspace*{-15pt} \\
 {\small $^{1}$D\'epartement de Physique, LIRST,  Facult\'e
Polydisciplinaire,
Universit\'e Sultan Moulay Slimane}\\
{\small B\'eni Mellal, Morocco}\\{\small $^{2}$ D\'{e}partement de
Physique, LASIMO, Facult\'{e} des Sciences, Universit\'{e} Ibn
Tofail,
 K\'{e}nitra, Morocco}  }   \date{}\maketitle
\begin{abstract}
We  study  four dimensional  quiver gauge models from F-theory
compactified on fourfolds with  hyper-K\"{a}hler structure. Using
intersecting complex toric surfaces,  we derive a class of  $N=1$
quivers with charged fundamental matter  placed on external  nodes.
The emphasis is on how local Calabi-Yau equations  solve  the
corresponding physical constraints including the anomaly cancelation
condition. Concretely, a linear chain of SU($N$) groups with flavor
symmetries has been constructed using polyvalent toric geometry.
 \\
 \\
\textbf{Keywords}: {\it F-theory; D7-branes  in Type IIB
superstring; Quiver gauge theories; Toric geometry.}
\end{abstract}

\newpage
\section{Introduction}
Study of  supersymmetric gauge theories, in various dimensions, has
attracted much attention. Concretely, they have been extensively
studied in the context of  II superstrings, M-theory,  and F-theory
using different  technics. The main reason  behind  such an interest
is that the relevant physics can be derived, nicely,  from geometric
and topologic data of the internal manifolds [1-7]. Very often in
Calabi-Yau  manifolds fibrations, the gauge group and matter content
of the resulting models are associated with the singularities of the
K3 surface fibers and the non-trivial  base geometry respectively
using the so-called {\it quiver method} \cite{8}.   The latter  can
be used to encode the field content  of a gauge theory  on a graph
(quiver). Precisely,  each node  of the quiver is assigned a gauge
group  factor and to each link a matter, as the bi-fundamental and
fundamental representations, charged  under the local gauge
symmetry. This approach has played a fundamental r\^ole in the
interface of higher dimensional physics and geometry.

 Recently,  there
have been many interesting developments  in F-theory using geometric
engineering method  exploring either  toric geometry  or   mirror
symmetry, to deal with  Calabi-Yau singularities.  A special
interest has been devoted to six-dimensional models from F-theory on
an elliptically fibered Calabi-Yau threefold using the connection
with ADE singularities based on  Lie algebras \cite{9}. It has been
suggested that many four dimensional  theories could  be derived
from a single
 six dimensional  origin \cite{90}.

 A close inspection, in the study of quivers,  shows that the
 fundamental piece used in  the deformation of the
Calabi-Yau singularities  is a collection of intersecting complex
curves. Generally,  they are identified with     one complex
dimensional projectile spaces   ($\bf CP^1$'s)\cite{1,2}.  The
corresponding intersection theory is intimately related  to the
Cartan matrices of Lie algebras. In this way, the Euler
characteristic  of toric manifold $\bf CP^1$ plays a more
established r\^ole due to its connection  with the  diagonal entries
of such  matrices.  In string theory,  this connection has provided
many quiver families with interesting  physical properties. However,
it  is  quite natural to think about quiver gauge theories based on
other toric manifolds.

The  purpose of   this work  is  to contribute to such a geometric
program by
 elaborating   a class of  of $N=1$  quivers  in four dimensions obtained from
  intersecting   two dimensional  toric manifolds.
This suggests a connection with   F-theory compactification
providing a new family of quivers that deserves a deep study.  It
may give a new meaning of  the  intersection of  higher dimensional
toric  geometries encoding many physical information hidden in local
Calabi-Yau manifold compactifications.  More precisely, we  discuss
a class of quivers  from F-theory compactified on four complex
dimensional hyper-K\"{a}hler manifolds  $X^4$. The manifolds are
defined as the moduli space of  $N=4$ sigma model in two dimensions.
Analyzing the corresponding  physical constraints,  we  reveal that
these manifolds  can be identified with    the cotangent fiber
bundle over intersecting 2-dimensional complex toric varieties
$V^2$. Compactifying F-theory on these  geometries,  we obtain a
class of $N=1$  quivers based on intersecting $V^2$'s represented by
polyvalent nodes. We show that the anomaly cancelation constraints
can be solved using toric geometry and intersection theory of local
Calabi-Yau manifolds.

 This  paper is organized as follows. Section 2
gives a consice review on F-theory. The compactification of F-theory
on four complex dimensional hyper-K\"{a}hler manifolds  $X^4$ is
presented in section 3. The construction  of the corresponding $N=1$
quivers in four dimensions using toric geometry method is given  in
section 4. A possible generalization based on polyvalent geometry,
producing a linear chain of SU($N$)  groups  with  a flavor
symmetry, is  proposed  in section 5. The last section is devoted to
concluding remarks.

\section{F-theory}
In this section we give a quick review of the relevant geometric
backgrounds of F-theory.  Following Vafa \cite{10},  this theory
describes a non-perturbative vacuum of Type IIB superstring in which
the dilaton and axion moduli
 are considered  as dynamical ones. This  has been supported by the implementation
 of a complex scalar field  interpreted as the complex structure
  parameter $\tau$ of
 an elliptic curve  providing, with Type IIB superstring, a
twelve-dimensional space-time.  In this way, Type IIB superstring
theory   can be seen as  the compactification of F-theory on the
elliptic curve $T^2$, by ignoring  the  size parameter contribution.
Considering F-theory, one can build lower dimensional models by
exploring the compactification on elliptically fibered Calabi-Yau
manifolds\cite{11}. The famous example playing a crucial r\^ole, in
such activities,  is F-theory on elliptically fibred K3 manifolds.

 Geometrically,  one
 build such fiber  manifolds  by varying the complex parameter  $\tau$ over the
dimensional projective space $\bf CP^1$ parameterized by the local
coordinate $z$. In this way, $\tau$ becomes a  function of $z$ as it
varies over the $\bf CP^1$ base of such  elliptic K3 surfaces. This
class of manifolds  can be  obtained  by considering    complex
surfaces given by the following equations
\begin{equation}
 y^2 = x^3 + f(z)x + g(z)
\end{equation} where
$f(z)$ and $g(z)$ are polynomials of degree 8 and 12, respectively.
The homogeneity condition is required by the triviality of the
canonical line bundle  ensuring the Calabi-Yau  condition. This
equation has been extensively studied in many places in connection
with elliptic fibration of F-theory \cite{6,7}.   It is shown that
these manifolds  have  24
 singular points associated with  $\tau(z) \to \infty$.  These singular points  are associated with the vanishing
 condition of  the following discriminant

\begin{equation}
 \Delta= 4f^3(z)x +27 g
 ^2(z).
\end{equation}
 In  D-brane physics,
 these singularities have been  identified with   D7-brane locations  producing  interesting gauge
 theories in eight dimensions\cite{10,11}. However, the connection with real world requires four
 complex dimensional manifolds in the compactification scenario.  In
 fact, one can consider  F-theory on elliptic K3 fibration over  a complex surface $S$.
Indeed,  $N=1$ supersymmetry in four dimensions  can be obtained by
imposing some geometric constraints on  $S$. It has been shown that
$S$  should satisfy the following Hodge number condition
\begin{equation}
\label{c} h^{1,0}= h^{2,0}=0.
\end{equation}
In \cite{6,7}, it has been   reveled that these two conditions kill
the adjoint chiral fields associated with Wilson lines and the zero
mode of the canonical line bundle producing models with only four
charges in four dimensions.

 In what follows, such surfaces will be
identified with two dimensional toric manifolds, which will be noted
by $V^2$. The corresponding intersection theory will be explored  to
build a class of $N=1$ quivers with flavor symmetries.

\section{    F-theory geometric  backgrounds }
  In this section, we discuss   F-theory on four complex
dimensional manifolds  $X^4$ associated with  the following
compactification
\begin{equation}
 {\mathbb{R}^{1,3}}\times X^4.
\end{equation}
Mimicking the analysis used on the geometric engender of $N=2$, we
consider a class of   manifolds satisfying the physical condition
given in eq.(\ref{c}). The  manifolds  will be identified with local
backgrounds considered as  the cotangent  fiber bundle over
intersecting two dimensional toric varieties $V^2$. This may offer
as  a possible generalization of intersecting  one dimensional
projective spaces $\bf CP^1$'s appearing  in local K3 surfaces.  It
is recalled that $V^2$ can be represented by toric diagrams
(polytopes) $\Delta(V_2 )$ spanned $2+r$ vertices $v_i$ belong to
the  $Z^2$ lattice \cite{12,13,14}. These vertices verify toric
constraints given by
\begin{equation}
\sum\limits_{i=1}^{2+r}q_i^av_i=0, \qquad a=1,\ldots,r,
\end{equation}
where $q^a_i$   are the corresponding Mori vectors.  Probably, the
most  simple example is ${\bf CP}^2$ defined by $r = 1$ and the Mori
vector charge $q_i = (1, 1, 1)$. The corresponding polytope  has 3
vertices $v_1$, $v_2$ and $v_3$  of the $Z^2$ square lattice
satisfying the  toric relation
\begin{equation}\label{cp2}
v_1+v_2+v_3=0.
\end{equation}
 This defines a triangle $(v_1v_2v_3)$, described by the intersection of
three ${\bf CP}^1$ curves,  in the $R^2$ plane.

Roughly speaking,  F-theory local geometries that we  will  examine
here can be  constructed   using technics of $N=4$ sigma model in
two dimensions \cite{140,15,16}.  Precisely, they  are interpreted
as the moduli space of two dimensional  $N=4$ supersymmetric  field
theory with $\mbox{U}(1)^r$ and  $(r+2)$ hyper-multiplets  assuring
the right dimension on which F-theory  should   be compactified. In
this way, the manifolds are obtained by solving   the following
D-flatness conditions
\begin{equation}\label{sigma4}
\sum_{i=1}^{r+2}q_{i}^{a}[\phi _{i}^{\alpha }{\bar{\phi}}_{i\beta
}+\phi _{i}^{\beta }{\bar{\phi}}_{i\alpha
}]=\vec{\xi}_{a}\vec{\sigma}_{\beta }^{\alpha },\;\;a=1,\ldots ,r
\end{equation}%
Here,   $q_{i}^{a}$ describe now   the sigma model  matrix charges.
For each hypermultiplet, $\phi _{i}^{\alpha }$'s ($\alpha =1,2)$ are
the component field doublets.  It is  noted that $\vec{\xi}_{a}$ are
interpreted  as   the Fayet-Iliopolos (FI) 3-vector couplings
related  by $\mbox{SU}(2)$ symmetry.  $\vec{\sigma}_{\beta }^{\alpha
}$ are the traceless $2\times 2$ Pauli matrices.   It is remarked
that the solutions of eqs(\ref{sigma4}) depend on many physical data
including  the number of the gauge fields, charges and the values of
the FI couplings.  Using   SU(2) R-symmetry transformations $\phi
^{\alpha }=\varepsilon ^{\alpha \beta }\phi _{\beta
},\;\overline{\phi ^{\alpha }}=\overline{\phi }_{\alpha
},\;\varepsilon _{12}=\varepsilon ^{21}=1$ and exploring  the Pauli
representation, eqs.(\ref{sigma4}) can be split as
follows\begin{equation} \label{2.2}
\begin{array}{lcr}
\sum\limits_{i=1}^{r+2}q_{i}^{a}(|\phi _{i}^{1}|^{2}-|\phi
_{i}^{2}|^{2})
&=\xi _{a}^{3}    \\
\sum\limits_{i=1}^{r+2}q_{i}^{a}\phi _{i}^{1}\overline{\phi
}_{i}^{2} &=\xi
_{a}^{1}+i{\xi ^{2}}_{a} \\
\sum\limits_{i=1}^{r+2}q_{i}^{a}\phi _{i}^{2}\overline{\phi
}_{i}^{1} &=\xi _{a}^{1}-i{\xi ^{2}}_{a}.
\end{array}
\end{equation}
 Since   eqs.(\ref{2.2})  is invariant under   the  U(1)$^r$ gauge transformations,
  we   can deduce  precisely an
eight-dimensional toric hyper-K\"{a}hler manifolds, which will be
considered as  F-theory backgrounds. However,
 explicit  solutions of these geometries  depend on the values of the FI
   couplings.  Taking   $\xi^1_a=\xi^2_a=0$  and
  $\xi^3_a >0$,  eqs.(\ref{2.2}) describe the
  cotangent  fiber bundle over  complex  two-dimensional toric varieties.
  To understand these solutions in some detail, we consider the
  case associated with the
  cotangent fiber  bundle  $T^*( \bf CP^2$). The manifold is defined as the
 moduli space of $2D $ $ N =4$ supersymmetric U(1) gauge theory with one isotriplet
FI coupling  parameter $\vec{\xi}=(\xi^1,\xi^2,\xi^3)$ and  only
three hypermultiplets of charges $q^i_a=q^i=1$, where  $i=1,2,3$. In
this case, eqs.(\ref{2.2}) can be reduced be
\begin{equation}
\label{5}
\begin{array}{lcr}
\sum\limits_{i=1}^3( |\phi^1_i|^2-|\phi^2_i|^2) &= \xi^3 \qquad &
\\
\sum\limits_{i=1}^3 \phi^1_i
\overline{\phi}_{i2}&=\xi^1+i{\xi^2}\qquad &
\\
\sum\limits_{i=1}^3 \phi^2_i
\overline{\phi}_{i1}&=\xi^1-i{\xi^2}.\qquad &
\end{array}
\end{equation}
It is noted that,   for $\xi^1=\xi^2=\xi^3=0$, the  moduli space has
an $\mbox{SU}(3)\times\mbox{SU}(2)_R$ symmetry. It can be considered
as  a   cone over a seven manifold given by
\begin{equation}
\label{7} \sum_{i=1}^3(\varphi_{\alpha i}\overline{\varphi}_i^\beta-
\varphi_i^\beta\overline{\varphi}_{\alpha i})= \delta{ _\alpha
^\beta}.
\end{equation}
However,  the case $\vec{\xi}\neq\vec{0}$ shows tat the
$\mbox{SU}(3)\times\mbox{SU}(2)_R$ symmetry is explicitly broken
down to $\mbox{SU}(3)\times\mbox{U}(1)_R$.  When   $\xi^1=\xi^2=0$
and $\xi^3$ positive definite,  we recover the  cotangent fiber
bundle  over  $\bf CP^2$.  For  ${\phi}_{i}^2=0$, one  finds
\begin{equation}
 |\phi^1_1|^2+|\phi^1_2|^2+|\phi^1_3|^2 = \xi^3,
\end{equation}
defining    now the
 $\bf CP^2$  projective space, in $N=2$ sigma model langauge \cite{17}. On the other hand,  with $\xi^1=\xi^2=0$
 conditions, the two last equations  of (\ref{5})  mean that
$ \overline{\phi}_{2i}$ lies in the cotangent space to  $\bf CP^2$.
This can be viewed as  an extension of the canonical  line bundle
over $\bf CP^2$ used in the study of $\mathcal{N}=1$  quivers
embedded in type II superstrings.   Putting $\xi_i^2=\xi_i^3=0$ and
introducing
 $x_i=\phi^1_i$  and $y_i=\phi^2_i$, the  two last  equations
 (\ref{2.2}) can be rewritten as
 \bea
\label{extendedA} \sum_{i=1}^3 x_i\bar{y_i}&=&0 \nonumber\\
\sum_{i=1}^3\bar{x_i}y_i &=& 0,  \eea  interpreted as  two
orthogonal variables. Then, the total space is a cotangent fiber
bundle over $\bf CP^2$. Locally, it can be identified with
\begin{equation}
C^2\times V^2
\end{equation}
To connect this geometry with the elliptic curve  fibration, we
orbifold the fiber $C^2$ by a by subgroups $\Gamma$ of SU(2) to
build spaces  of the form  $ {\bf C}^2/\Gamma$. This can produce a
fibration of an elliptic curve over the plan
\begin{equation}
{\bf C}/\Gamma\times {\bf C}
\end{equation}
 In fact, the elliptic curve ${\bf C}/\Gamma$ can be controlled by
 eq.(1). This  can produce  local K3 surfaces  associated with  D7-brane locations  in  Type IIB
superstring.  The construction of such manifolds goes back to many
years ago  and it has been explored to elaborate $ADE$ gauge
symmetries in eight dimensions \cite{10,11}.

Motivated by Standard Model (SM) and its extensions, it should be
interesting to consider   four dimensional quiver gauge theories
from such geometries. In fact, this can be obtained by considering
intersecting two dimensional complex  toric $V^2$'s.  Indeed, the
previous example can be generalized  by considering $\mbox{U}(1)^r$
gauge theory in $N=4$ sigma model.  This can be analyzed to show
that eqs.(\ref{sigma4}) describe the cotangent fiber  bundle over
$r$ intersecting $V^2_a$  ($a=1,\ldots,r$).  These geometries can
arise as a generalization of  intersecting $\bf CP^1$'s of the  K3
surfaces which are classified by Lie algebras \cite{10,11}.

In  the  present geometry,  $V^2_{a}$ form  a basis of the  middle
cohomology group $H_4(X^4,Z)$. The  corresponding intersection
numbers are given by  the square matrix
\begin{equation}
[V_a^2].[V^2_b]=I_{ab}.
\end{equation}
Toric geometry assures that $V^2_{a}$  intersects $V^2_{a+1}$ at a
single  $\bf CP^1$   which  produces the following intersection
numbers
\begin{equation}
[V^2_{a}]\cdot \lbrack V^2_{a+1}]=-2.
\end{equation}
Assuming that $V^2_{a}$ does not intersect  $V^2_{b}$ if $|b-a|>1$.
As in the case of ${\bf CP}^1$'s, the intersection  matrix  should
take the following form
\begin{equation}
I_{ab}=n\delta_{a\;b}-2(\delta_{a\;b-1}+\delta_{a\;b+1}),
\end{equation}
 where  $n$ is  the Euler characteristic of $V^2$, noted   usually
 by
 $\chi(V^2)$.

Having constructed   a special type  of fourfolds  manifolds. The
next section will be concerned with  the associated quiver.
Concretely, we will discuss the connection between the physics
content of F-theory on such  manifolds and quiver method of $N = $?
supersymmetric gauge theories in four dimensions. The analysis that
will be used   here is  based on  polyvalent toric  geometry.

\section{Quiver gauge theories from F-theory}
To start, it is  recalled that four dimensional gauge theories has
attracted a special attention in connection with string theory on
Calabi-Yau manifolds. Concretely, they   have been investigated in
the study of D-brane
 gauge theories obtained  from  the  compactification on
singular  Calabi-Yau manifolds or $G_2$ manifolds \cite{4}. In this
way,  the   gauge symmetry  and  the matter content of the resulting
models  can be derived  from the geometric and topological data  of
the internal manifolds[1-7].  The  associated physical parameters
are  related to theirs moduli spaces. A nice way to encode the
physical information of such gauge theories is to use the quiver
method\cite{8}. The corresponding field  models are usually refereed
to  quiver gauge theories.  In this way, The physical content of  a
model with several gauge group factors
\begin{equation}
 \label{G}
 G =\prod_{a}G_a,
\end{equation}
can be illustrated    by a quiver. As  in graph theory,  it is
formed by nodes and links.  For each  node, one associates a gauge
factor $G_a$, where $a$ indicates  the number of nodes. Here, it can
be identified with the number  of $V_a$. However, links between two
nodes are associated with charged matter transforming either in
bi-fundamental or fundamental representations of the gauge group
$G$.

In string theory, many examples of  such quivers have been
elaborated  in connection with toric graphs and Dynkin  diagrams. It
is worth noting that  the   last  class  is the most studied one due
to its link   with  D-brane gauge theories obtained from Calabi-Yau
singularities classified by  Cartan matrices of Lie algebras. This
has  led to a nice classification of stringy  quivers. Details on
the corresponding materials can be found in, for example,
\cite{1,2}. It is remarked that these classes  have been based on
the intersection of $\bf CP^1$'s.

Using higher dimensional intersecting geometries, we would like to
engineer $N=1$  quivers in  the  F-theory contexts using Type IIB
superstring dual descriptions in terms of D7-branes. It is recalled
that F-theory
 contains also  D3 and D5-branes localized at some singular
points. However, to understand the present construction, we focus
for simplicity,  on  D7-branes wrapping  two complex dimensional
toric varieties  and filling  the four-dimensional Minkowski space.
Roughly speaking,  it has been shown that  F-theory near the
$A_{N-1}$ singularity of the K3 surfaces is equivalent to $N$ units
of D7-branes in Type IIB dual producing SU($N$)   gauge theories  in
eight
 dimensions \cite{10,11}.   Four dimensional quiver gauge theories
 can be  reached by wrapping such  D7-branes on  intersecting  $V^2_a$. Indeed,
 consider   $r$ different stacks of D7-branes.  By standard results in quiver gauge
 theories,
  the  associated  gauge group reads as
\begin{equation}
\label{G}
 G\ =\ \bigotimes_{a=1}^r SU(N_a), \qquad N=\sum_a N_a
\end{equation}
where   $N_a$   are  integers  which characterize  the   physical
content. In general, these integers   are determined by imposing
physical requirements on the internal manifolds.  In the presence of
fundamental matter, the  anomaly cancelation
    condition   for the theory with only  four supercharges  responds to
  the  following constraint equations
\begin{equation}
\label{intersection}
 \sum_{a=1}^rI_{ab}N_a -M_b=0,
\end{equation}
where $I_{ab}$   is the  intersection  matrix  of  $V^2$'s  on which
D7-branes   are  wrapped.   $M_b$  are fundamental matter fields,
charged under the gauge group $G$. To get the corresponding quivers,
one should solve these physical constraints. A priori,  they are
many ways to solve them. However, a close inspection shows that it
can be interpreted as  a toric geometry realization of local
Calabi-Yau manifolds \cite{2}. To understand this way in some
details, we present explicit models. For example,  we consider two
gauge factors ${\mbox{U}(N_1)}\times {\mbox{U}(N_2)}$  quiver gauge
theory associated with two intersecting
 generic toric manifold $V^2$'s  at $\bf CP^1$.  The  main simplifying
 feature of this geometry is that its intersection matrix  reads as
 \begin{equation}
I_{ab}=\begin{pmatrix}
n & -2 \\
-2 & n
\end{pmatrix}.
\end{equation}
To solve  the associated  physical  equations (\ref{intersection}),
one may implement auxiliary nodes producing a new quiver  with more
than two gauge factors. We note   that this geometric  procedure
allows one to modify   the above intersection matrix  leading to a
Calabi-Yau geometry without affecting  the dynamical  gauge factors.
Indeed, to get the desired quiver the extra gauge factors   behave
like non dynamic ones and they should be associated with fundamental
matter fields. In geometric engineering method,  this can be
obtained by assuming that auxiliary nodes  correspond to D7-branes
wrapping in non compact cycles. Geometrically, one should consider
the following modified intersection  matrix
\begin{equation}
\widetilde{{I}}_{ai}=\left(
  \begin{array}{cccc}
   2-n& n&   -2& 0 \\
    0 & -2&n&2-n \\
  \end{array}
\right) \label{matrix4}
\end{equation}
as required by   the local Calab-Yau condition

 \be
  \sum_{i=1}^4 \widetilde{{I}}_{ai} =0, \qquad a=1,2.
  \label{lcy}
  \ee
 The associated toric geometry is defined by a polyhedron which is
a convex hull of   four vertices   \be
  v_i=(v^1_i,v^2_i,v^3_i),\qquad i=1,\ldots,4
 \label{toric}
 \ee
satisfying the  following toric relations \be
 \sum_{i=1}^4 \widetilde{{I}}_{ai} v_i=0, \qquad a=1,2.
 \label{toric}
 \ee
In fact, these  equations look like a generalization of
(\ref{intersection}).  Up to some details,  the corresponding
physical quantities, like  the number of fundamental matter and the
gauge group ranks, can be identified with toric data including the
Mori vector charges. One can obviously see that  one  of the vertex
entries should satisfy the following requirement \bea  v^1_1=M_1
\nonumber\\
 v^1_2=N_1
\nonumber\\ v^1_3=N_2
\\ v^1_4=M_2
\nonumber\eea   In turn, the  intersection
 matrix  can be  regarded as   Mori  vectors and take the form
 \be
I_{ab}=\widetilde{ {I}}_{ab}, \qquad a=1,2.
 \label{toric}
 \ee
It follows from toric geometry that the equation
(\ref{intersection})  can be solved by \bea
  N_1&=&N_2=N
\nonumber\\
   M_1&=&M_2=(n-2)N.
  \eea
  This solution  can produce a quiver with fundamental   charged
  matter which   can  be obtained  by  zero  limits of the
gauge coupling constants associated with two auxiliary  nodes. In
toric geometry, they correspond to   the extra lines and columns of
the triangular matrix (\ref{matrix4}). In practice, this can done by
considering cycles with  very large volume.  In this way, the
dynamics associated with such cycles become week and lead to  a
spectator flavor symmetry. This assumption provides  a quiver theory
with U$(N)\times$U$(N)$ gauge symmetry and flavor group  of type
U$(nN-2N)\times$U$(nN-2N)$ placed  non physical nodes. The quiver
illustrating  the  physics content  is shown in figure 1. In this
quiver, external nodes represent gauge  symmetry factors, while  the
external ones  indicate flavor factors.

\begin{figure}[h]
        \begin{center}
            \includegraphics[scale=0.5]{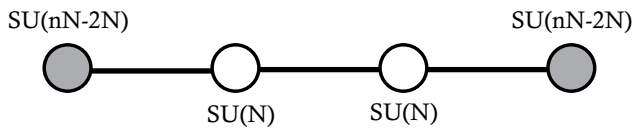}
        \end{center}
        \caption{Bivalent quiver. }
\end{figure}

It is interesting to note this model is based on bivalent geometry.
The analysis may be extended to quivers involving nodes  with more
than two links. This introduces polyvalent vertices in  toric
geometry representation of local  Calabi-Yau manifolds.   We discuss
this issue in the following sections.

\section{Quivers from trivalent geometry}
In the engineering   method used  in Type II superstrings, trivalent
geometry however contains both bivalent and trivalent nodes. It
contain central nodes linked with three other  ones.  In the present
F-theory backgrounds, a central node will correspond to a Mori
vector charge of the form \be (n, -2, -2, 4-n, 0, \ldots, 0), \ee
 required by the local Calabi-Yau condition. To get the
quiver with such a geometry, one may follow the approach used for
the bivalent one. Indeed, we consider the geometry associated with
$\mbox{U}(1)^r$ gauge symmetry in $N=4$ sigma model. It is observed
that the trivalent geometry appears for $ r\geq 3$. To see how this
works,  we  first deal with a concrete example for $r=3$. The local
Calabi-Yau condition imposes the following modified intersection
matrix
\begin{equation}
\widetilde{{I}}_{ai}=\left(
  \begin{array}{cccccc}
   2-n& n&   -2& 0& 0&0\\
    0 & -2&n&-2&4-n&0 \\
     0& 0 & -2&n& 0& 2-n
  \end{array}
\right). \label{matrix6}
\end{equation}
 In order to construct the corresponding quiver,
the equation (\ref{intersection})  should be solved.  As the
previous model, the relevant physical data can be easily computed
using toric geometry. We, then, find  \bea
  N_1&=&N_2= N_3=N
\nonumber\\
  M_1&=&M_3=(n-2)N, \quad  M_2=(n-4)N.
  \eea
This solution indicates that ones expects  a quiver theory with
U$(N)^3$ gauge symmetry  and a  flavor  group  of type
U$(nN-2N)^2\times$U$(nN-4N)$ associated with non physical nodes.
This can  be represented  by a graph as  shown in figure 2.
\begin{figure}[h]
        \begin{center}
            \includegraphics[scale=0.5]{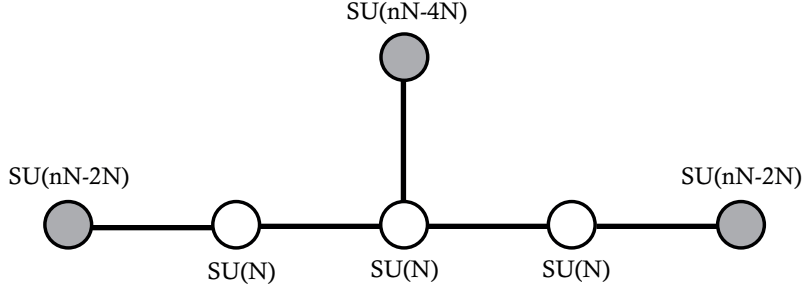}
        \end{center}
        \caption{Trivalent quiver.}
\end{figure}

The general  cases can be treated very similarly. It is easy to
extend the above matrices to a class of models  controlled by the
following  modified intersection matrix
  \bea
  \widetilde{ {I}}_{00}&=& \widetilde{ {I}}_{r+1 r+1}=(2-n),
\nonumber\\
 \widetilde{ {I}}_{ab}&=&I_{ab}, \qquad a=1,\ldots, r\\
\widetilde{ {I}}_{aa*2}&=&(4-n), \qquad a=1,\ldots, r-1 \nonumber
 \label{toric}
 \eea
with other vanishing. Using toric geometry,  it  is possible to
solve the equation (\ref{intersection}). As the previous cases, the
associated quiver can  be obtained  by zero limits of the gauge
coupling constants corresponding to  two bivalent  and  $r-2$
trivalent external nodes. This can produce a quiver theory with a
U$(N)^r$ gauge symmetry and flavor group  of type
U$(nN-4N)^{r-2}\times$U$(nN-2N)^2$ associated with the  spectator
nodes. In graph theory, this model  can be  represented  by the
quiver given in figure 3.
\begin{figure}[h]
        \begin{center}
            \includegraphics[scale=0.5]{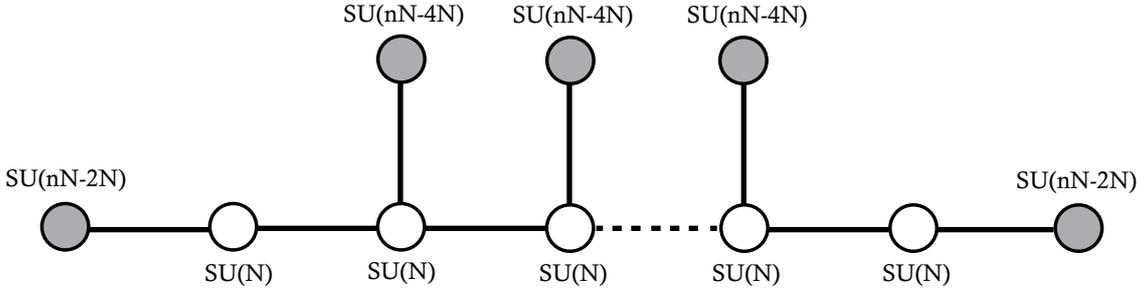}
        \end{center}
        \caption{Extended trivalent quiver.}
\end{figure}
\section{ Polyvalent geometry and quivers}
In this section, we   construct  quivers  involving polyvalent nodes
where the bivalent and trivalent ones appear just as the leading
objects  of a more general  picture.    This geometry  can be
considered as a possible extension  to the previous cases dealing
with bivalent and trivalent  quivers.  It is recalled  that a
polyvalent node can represent a central  toric variety $V^2$  with
self intersection $n$ intersecting $m$  other ones with contribution
(-2). Like for the bivalent and trivalent case,  the  local
Calabi-Yu condition impose the implementation of an  extra non
compact toric variety  with contribution $2m-n$. This requires that
F-theory quiver geometry should  involve the following Mori vector

 \be ( n, \underbrace{-2, -2, \ldots, -2}_{m}\;, 2m-n, 0\ldots, 0).\ee
 Instead of being general, let us consider a concrete  example. Probably,
 the most simple one is a  quiver involving just one
 tetravalent  node. In geometric engineering  method,   this has  been  considered as
  is a particular example of polyvalent
nodes  coming  after the bivalent and trivalent  ones. In connection
with Dynkin diagrams,  the  tetravalent geometry appears in the
affine so(8) Dynkin diagram. To understand  that, we illustrate the
tetravalent geometry.  The  quiver corresponds to 4 intersecting
 generic toric manifold $V^2$'s  at $\bf CP^1$ controlled  by the
 following matrix
 \begin{equation}
I_{ab}=\begin{pmatrix}
n & -2 & 0&0\\
-2 & n&-2&-2\\
0 & -2&n&0 \\
0&-2 & 0&n
\end{pmatrix}.
\end{equation}
 Local Calabi-Yau condition (\ref{intersection}) requires the introduction   of
    an auxiliary node producing a   tetravalent  geometry.  The latter
 correspond to the following Mori vector
 \be (n, -2, -2, -2, 6-n, 0, 0,0, 0). \ee
  To solve the  previous physical conditions,  toric  geometry produces the following equations \be
 \sum_{i=1}^8 \widetilde{{I}}_{ai} v_i=0, \qquad a=1,2,3,4
 \label{toric}
 \ee
 with the local  Calabi-Yau condition

  \be
  \sum_{i=1}^8 \widetilde{{I}}_{ai} =0, \qquad a=1,2,3,4
  \label{lcy}
  \ee

 Similar calculations  can be used to build a  quiver with fundamental   charged
  matter  placed on  a tetravalent and 3 bivalent nodes as  shown in figure 4.
  \begin{figure}[h]
        \begin{center}
            \includegraphics[scale=0.5]{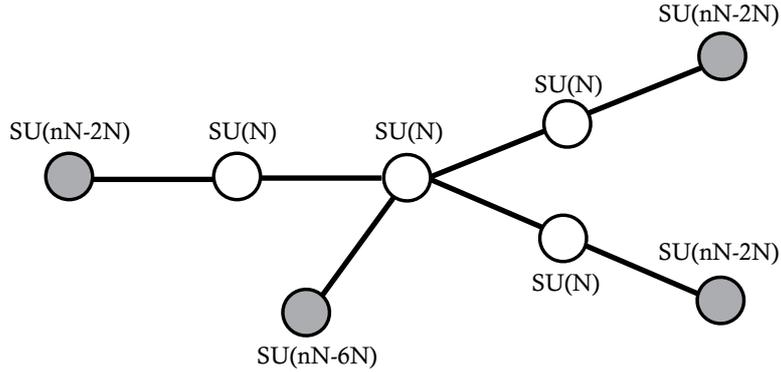}
        \end{center}
        \caption{Tetravalent  quiver.}
\end{figure}

  This  geometry  provides  a quiver theory
with U$(N)^4$gauge symmetry and flavor group  of type
U$(nN-2N)^3\times$U$(nN-6N)$.

\section{Discussions}
In this work, we  have investigated    a class of  of $N=1$  quivers
in four dimensions obtained from
  intersecting   two dimensional  toric manifolds.
This suggests a connection with   F-theory compactification
providing a new family of quivers that deserves a deep study.  It
may give a new meaning of  the intersection of higher dimensional
geometries encoding many physical information hidden in local
Calabi-Yau manifold compactifications.  Concretely, we  have
engineered  a class of quivers  from F-theory compactified on four
complex dimensional hyper-K\"{a}hler manifolds $X^4$. The manifolds
are defined as the moduli space of  $N=4$ sigma model in two
dimensions. In particular,  we have remarked that these manifolds
can be identified with the cotangent fiber  bundle over intersecting
2-dimensional complex toric varieties $V^2$. Conpactifying F-theory
on these geometries,  we have derived a class of $N=1$ quivers based
on intersecting $V^2$'s. Then, we   have shown that the anomaly
cancelation constraints  can be solved using toric geometry  of
local Calabi-Yau manifolds.

Explicitly,  we have engineered  quivers based on    a linear chain
of SU($N$) groups with flavor symmetries  depending on polyvalent
nodes. It turns out that   each node $a$   represents a toric
variety $V^2_{a,m}$ which intersects $m$  other ones.  This node
encapsulates a SU($N_c$) gauge factor and SU($N_f$)  flavor
symmetry.  Calculation, in toric geometry,  shows that \bea
 N_c&=&N
\nonumber\\
   M_f&=&(n-2m-2)N.
  \eea
This work comes up with many   open questions.  Motivated  by
\cite{18}, it should be interesting to build the metric of such
varieties.   On the other hand,  it has been observed that  there is
a similarity   with    quivers based on  finite Lie algebras. The
approach could be adaptable to a broad variety of geometries
associated with other Lie algebras.  It would therefore be of
interest to try to extract physical information on such  geometries
to give a complete study. \\
\\

{\bf Acknowledgments}: The authors thank Z. Benslimane for
associated collaborations.

\end{document}